\documentclass[aps,prl,reprint]{revtex4-1}%
\usepackage{amsmath}%
\usepackage{amsfonts}%
\usepackage{amssymb}%
\usepackage{graphicx}
\usepackage{epstopdf}
\usepackage{natbib}

\begin{document}

\title{Cascading Failures in Networks with Proximate Dependent Nodes}
\author{Yosef Kornbluth, Steven Lowinger, Gabriel Cwilich, and Sergey V. Buldyrev}
\affiliation{Department of Physics, Yeshiva University, 500 West 185th Street, New
York, New York 10033, USA}

\date{\today}

\begin{abstract}
We study the mutual percolation of a system composed
of two interdependent random regular networks. We introduce a notion of distance to explore the effects of the proximity of interdependent nodes on the cascade of failures after an initial attack.  We find a non-trivial relation between the nature of the transition through which the networks disintegrate and the parameters of the system, which are the degree of the nodes and the maximum distance between interdependent nodes. We explain this relation by solving the problem analytically for the relevant set of cases. \end{abstract}
\maketitle
\begin{figure}
\includegraphics[clip,width=\columnwidth]{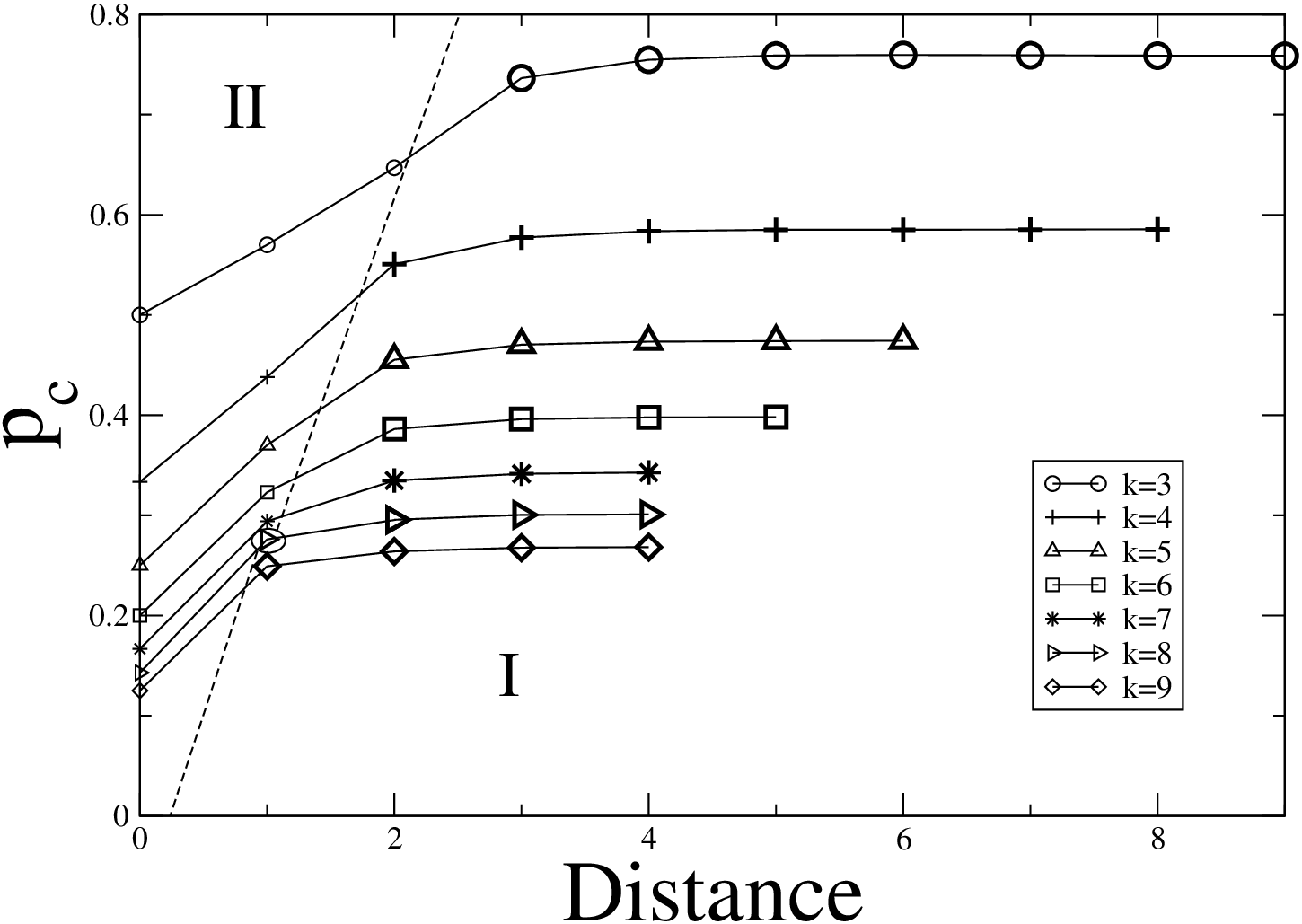}
\caption{Numerical results of $p_c$ vs. distance, obtained by averaging
  $100$ realizations of networks consisting of $N=10^6$ nodes for
  different degrees of connectivity ($k$). Large symbols denote first-order transitions, while second-order transitions are represented by small symbols. The regions of first-order and second-order transitions are denoted on the
graph and are separated by the dashed line. The ellipse around the value $p_c(8,1)$ indicates the special case of almost coinciding 
first- and second-order transitions.}
\label{f:pvk}
\end{figure}
\begin{figure}
\includegraphics[clip,width=\columnwidth]{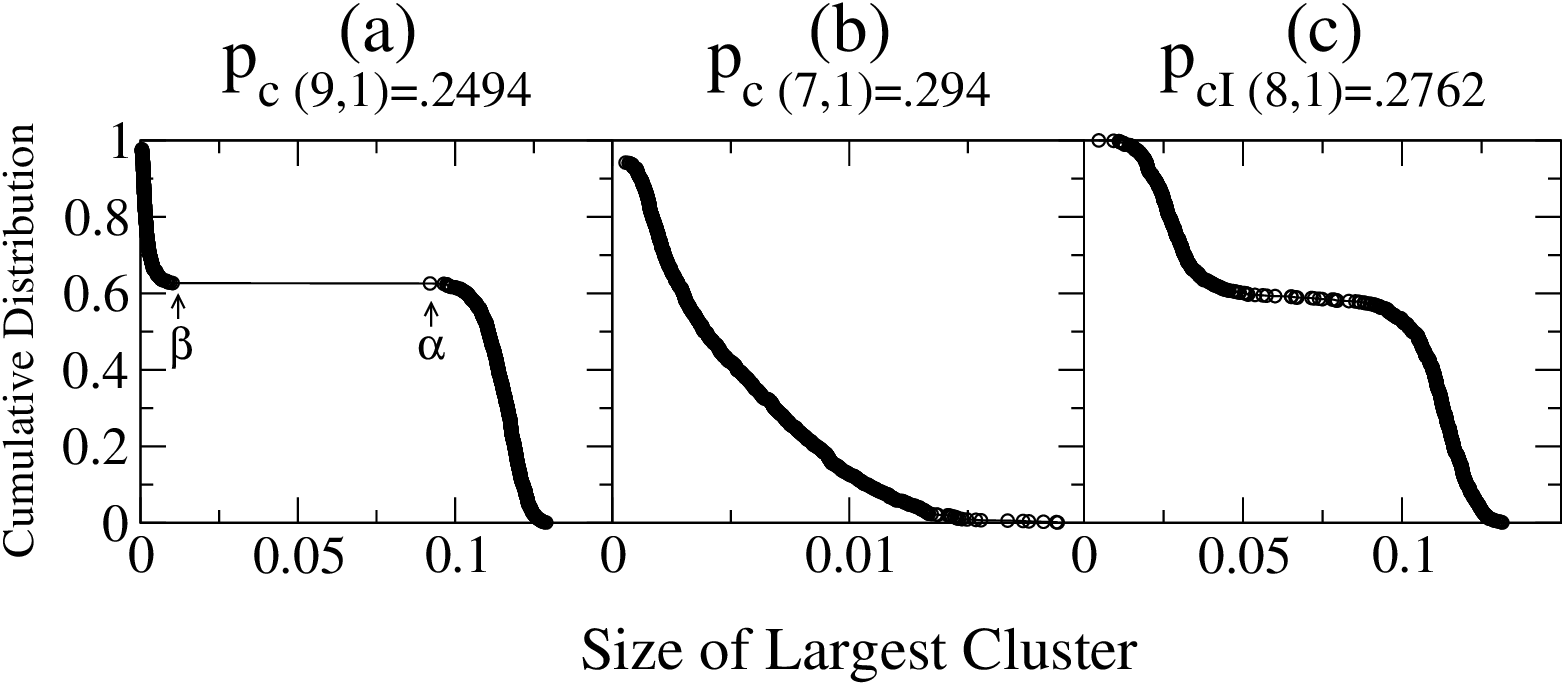} 
\caption{The cumulative distribution of the fraction, $\mu$, of nodes in the largest cluster at $p_c$. Each data point represents the largest cluster for
  1 of 1000 realizations. These graphs represent networks of $10^6$ nodes; similar patterns can be found for other sizes. In  panel (a), a first-order transition, none of the
  simulations result in $\beta << \mu
  << \alpha$. In contrast, a second-order transition, shown in panel
  (b), displays a distribution of $\mu$ with a continuous derivative. The third
  figure shows the transition at $p_c(8,1)$; it is similar to a
  first-order transition, but the gap between $\beta$ and $\alpha$ cannot be clearly
  identified.}
\label{cluster}
\end{figure}

Previous studies of the robustness of interdependent networks have
focused on networks in which there is no constraint on the distance
between the interdependent nodes 
\cite{Huang, Parshani, Shao, Buldyrev, Hu, Buldyrevnat, Dong, Gao,
  Son}. However, many dependency links in the
real world connect nearby nodes. For example, the international
network of seaports and the network of national highways form a
complex system. As seen recently from the effects of Hurricane
Sandy in New York City, if a seaport is damaged, the city that
depends on it will become isolated from the highway network due to the
lack of fuel. Similarly, a city without roads cannot supply a seaport properly. However, a city will depend on a nearby seaport, not
on one across the world.  Li et. al \cite{Li} investigated
distance-limited interdependent \textit{lattice} networks by computer simulations and found that allowing only local interdependency links changed the resilience properties of the system. Here, we study the analytically tractable random networks.\\
We study the mutual percolation of two interdependent
random regular (RR) graphs. We build two identical networks, A and B,
each of whose nodes are labeled $1...N$. Each node is randomly
connected by edges to exactly $k$ other nodes, in such a way that the two networks
have identical topologies. We then create one-to-one bidirectional dependency links, requiring that  
the shortest path between the interdependent nodes does not exceed an 
integer constant $\ell$.  \\
Formally, we establish two isomorphisms between 
networks A and B, a topological isomorphism and a dependency 
isomorphism. The topological isomorphism is defined for each node 
$A_i$ as $T(A_i)=B_i$ and $T(B_i)=A_i$. If $A_i$ and $A_j$
are immediate neighbors in network A, then $B_i$ and $B_j$ are
immediate neighbors in network B and vice versa.
The dependency isomorphism has the property that if $D(A_i)=B_k$, then $D(B_k)=A_i$. We create a restriction that $B_k=D(A_i)$ only if there are a maximum of
$\ell$ connectivity links on the shortest path connecting $A_i$ and
$A_k=T(B_k)$. \\
For simplicity, we introduce two more restrictions. We only set $D(A_i)=B_i$ if there are no other possibilities for $D(A_i)$. Additionally, we require that if $D(A_i)=B_k$, then $D(B_i)=A_k$. This restriction decreases the computation time needed.  Simulations suggest that the results for the models with and without the latter restriction are indistinguishable within the statistical error.\\
Following the mutual percolation model described in Buldyrev et
al. \cite{Buldyrevnat}, we destroy a fraction $(1-p)$ of randomly
selected nodes in A. Any nodes that, as a result, lost their connectivity links to the largest cluster (as defined in classical, single-network percolation theory
\cite{newmanbook,havlinbook}) are also destroyed. In the next stage, nodes in B that have their interdependent nodes in the other network destroyed are also destroyed. Consequently, the nodes that are isolated 
from the largest cluster in B 
as a result of the destruction of nodes in B are also destroyed. The iteration of this process, which alternates between the two networks, leads to a cascade of
failures. The cascade ends when no more nodes fail in either network.  The pair of remaining largest interdependent clusters in both networks is called a largest mutual component. If in the thermodynamic limit
$N\to\infty$, the fraction of nodes $\mu$ in the largest mutual component is greater than zero, it is called the giant mutual component.\\
As $p$
decreases, the giant component also decreases in size and eventually disappears at the value of $p$ that we call critical $p$, or $p_c$. We will denote $p_c$ for a
 degree $k$ and distance $\ell$ as $p_c(k,\ell)$. We investigate how $p_c$ varies as
a function of $k$ and $\ell$ (Figure \ref{f:pvk}).\\
As the distance and the degree increase, the transition at $p_c$ shifts
from a second-order transition in which $\mu$ continuously approach zero to a first-order one, in which $\mu$ decreases discontinuously in the thermodynamic limit. The line separating these two regimes is shown in Fig. \ref{f:pvk}. Of particular note is the shift
for $p_c(k,1)$. If $3 \leq k \leq 7$, the transition is second
order. For $k \geq 9$, the transition is first-order. For $k=8$, we
observe two transitions: the first-order transition at $p=p_{cI}=0.276$ followed by 
the second-order transition at $p=p_{c II}=0.269$. As we will explain, the existence of the second-order transition in this case affects the behavior of the first-order transition.\\
For each value of $k$, $p_c$ increases monotonically as $\ell$ increases. This monotonic increase is in contrast with the results
found by Li et al. \cite{Li}, who found that in distance-limited
lattices, there is a maximum in $p_c$ as a function of $\ell$. In Li et al.,
this maximum coincided with the distance at which the transition
shifts its nature from second-order for smaller distances to
first-order for larger distances. In our model, a similar shift occurs
at a very low value of $\ell\leq 2$ and is not associated with a maximum of
$p_c$.\\
If the transition is of the first-order,
the size of the largest mutual component of a finite network at a fixed value of $p\approx p_c$ may fluctuate dramatically. Due to
fluctuations always present in a finite system, $\mu$ can be close to the value $\alpha$, which is the fraction of nodes in the mutual giant component for $N\to \infty$ at $p=p_c+\epsilon$, where $\epsilon$ is a small positive number. In other cases, it can be close to $\beta$, the value of $\mu$  at $p=p_c-\epsilon$ for $N\to\infty$. However, no values of $\mu$ are expected in the interval between $\alpha$ and $\beta$. 
Fig. \ref{cluster} (a) illustrates the first-order transition observed at $p_c(9,1)=.2494$; we see that there is no
intermediate value of $\mu$ when the simulation is repeated for
many such systems. As the size of the network increases,
the distribution concentrates around $\alpha$ and $\beta$, with the probability density approaching a delta function at these two points. We define $p_c$ as the point where half of the realizations lead to each result. As the size of the network increases, $\beta$ goes to 0 in most cases. The sole exception is $p_{cI}(8,1)$ (Fig. \ref{cluster} (c)).\\
In the case of second-
order transitions, the drop in $\mu (p)$ is less dramatic. There is no
gap in the size distribution of the largest giant component (Fig. \ref{cluster} (b)). Rather, there is a steady decline of the cumulative distribution of $\mu$. Thus, $\alpha$ and $\beta$ are not defined. \\
These two categories describe every transition with the exception of $p_c(8,1)$. There, we see a plateau in the cumulative
distribution of the largest component size, reminiscent of a first-order transition. However, there is no clear gap 
in the distribution of the mutual giant 
component in a finite network and we cannot clearly identify $\alpha$ and $\beta$ (Fig. \ref{cluster}).
To explain the curious behavior of $p_{cI}(8,1)$, as well as to better understand the general behavior of interdependent networks with distance constraints, we offer a brief synopsis of our analytic solution for the case where $\ell=1$, which is the simplest case, in which we see the transition shift from second-order to first-order.\\
In our simulation, we establish dependency links in the following way:
At each step, we select at random a node $A_i$ that currently does not have a dependency link and set 
two dependency links $D(A_i)=B_j$ and $D(A_j)=B_i$, where $B_j$ is randomly chosen among $B_i$'s immediate neighbors without a dependency link, if at least one such neighbor exists. We call $(i,j)$ a dimer. If all of the neighbors of $B_i$ already have dependency links, we set $D(A_i)=B_i$ and $i$ is called a monomer. In order to calculate $p_c$, we find the fraction of indices that are monomers, 
\begin{equation}
m=\int_0^1\frac{2k e^{2\int_0^A\frac{(B^k-1)dB}{[B^k(k-2)-2(k-1)](B-1)}}}{2(k-1)-A^k(k-2)}dA-1.
\label{e:Renyi}
\end{equation}

Finding the value of $1-m$ is a version of R\'{e}nyi's parking
problem \cite{Renyi} that applies to a discrete graph. 
We will present our full analytic solution of that problem and the derivation of Eq. (\ref{e:Renyi}) elsewhere.\\ 

In our analytic calculation of $\mu(p)$ we will use the probability $q$ that a link leaving a dimer reaches a monomer. Because, in our algorithm, two monomers cannot be adjoining, any connectivity link that has a monomer at one end has a dimer at the other end. Thus, $q$ is simply
the number of links with monomers at one end divided by the number of
links that leave dimers, which is $\frac{km}{(k-1)(1-m)}$.\\
 
\begin{figure}
\includegraphics[clip,width=\columnwidth]{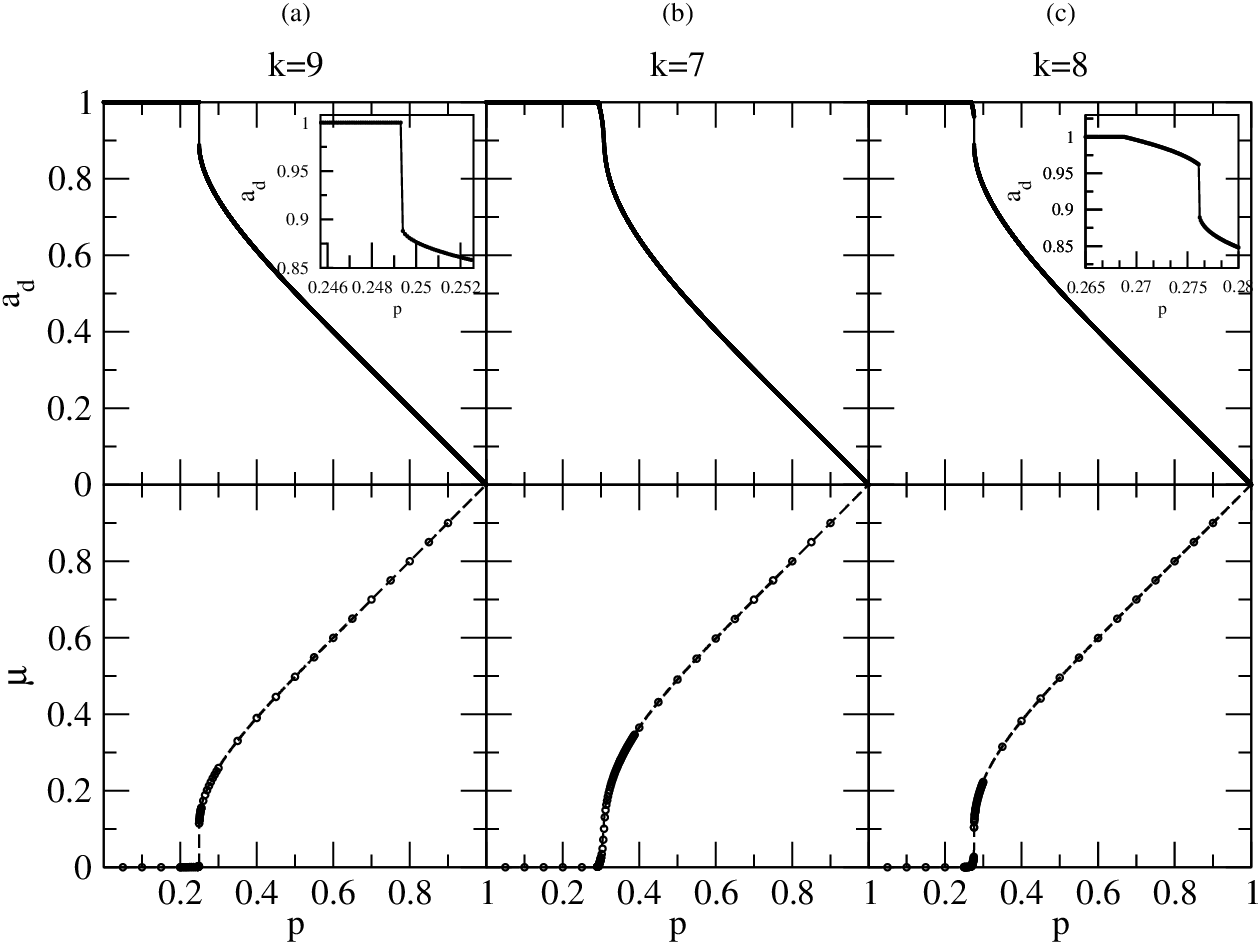}
\caption{The fraction of nodes in the mutual giant component $\mu(p)$ for $\ell=1$ and (a) $k=9$, (b) $k=7$ and $(c)k=8$. Dotted lines in the lower graph represent the
  fraction of nodes in the mutual giant component ($\mu$) derived from the analytic solution. The circles represent the values obtained
in simulation of networks with $N=10^6$ nodes, which show excellent agreement
with theory. The upper graph represents $a_d$, one of the probabilities defined in Eqs. (\ref{e:ad})-(\ref{e:zm}). The other probabilities behave similarly. Insets show the detailed behavior of $a_d(p)$ near the transition revealing, respectively, a discontinuous increase to 1 characteristic of the first order transition ($k=9$), a continuous increase to 1 characteristic of the second-order transition ($k=7$), and a discontinuous increase to .9616 at $p=p_{cI}=0.2762$ followed by a continuous increase to 1 at $p=p_{cII}=0.2688$, indicating the existence of two transitions ($k=8$). In all cases, $a_d=1$ indicates  the complete destruction of the mutual giant component, as shown in the lower graph.}
\label{f:solve}
\end{figure}

When $\ell=1$, it is possible to express the problem of finding the mutual
giant component of two networks as the problem of finding of 
a giant component in a single network, while taking into account the disruption
caused by the other network.
To
determine $p_c$ for a single RR graph, percolation theory introduces a probability $b$
that a randomly selected link does not lead to the giant
component. This probability satisfies the equation \cite{newmanbook}
\begin{equation}
b=(1-p)+p b^{k-1}.
\label{e:perc}
\end{equation}
The first term on the right side of the equation refers to the probability that the link leads to a node that was destroyed in the initial attack, while the second term represents the probability that the node survived, but does not have any outgoing links that lead to the giant component.
Previous studies of interdependent networks
have used this equation \cite{Son}, often modifying either the value
of $p$ \cite{Buldyrevnat, Shao, Parshani, Hu} or the degree
distribution \cite{Huang, Dong,
  Buldyrev}. They use these modifications to describe the diminished network at each stage of the cascade. However, this method is only useful when the nodes destroyed in each stage of the cascade can be seen as a random subset of nodes in the other network. The distance restriction in our model precludes us from doing so. We modify the equation itself to account for the local
interdependency between our two networks. This allows us to study both networks at once and to account for more complex relationships between them.\\
When studying a link in network A, we assume that the link
starts from a node $A_i$, such that its interdependent counterpart in network B, $B_k=D(A_i)$ {\it a priori} belongs to the mutual giant component. This assumption allows us to isolate the effect that the destruction of nodes in network A will have. Formally, this is done by assuming that $B_k$ is artificially connected to the mutual giant component in a way that does not affect the topology in network A. We thus define $a$ as the probability that a randomly chosen link leaving node $A_i$ would not lead to the mutual giant component, if $B_k=D(A_i)$ were to be artificially attached to the mutual giant component.
If the link leaving $A_i$ leads to a node $A_j$ that survived the initial
attack, $x$ represents the event that $A_j$ is not connected to the
mutual giant component via any of its outgoing links, even with our {\it a priori} assumption.
However, even if $x$ does not occur, $A_j$ may still be dead because its interdependent node
$B_l=D(A_j)$ does not belong to the mutual giant component. 
Accordingly, we define the event $y$ that the node $B_l$ is not
connected to the mutual giant component via links of network B. 
Thus, we can write
\begin{equation}
a=(1-p)+p[P(x \cup y)].
\label{e:gen}
\end{equation}
where $P(x \cup y)$ plays the role that
$b^{k-1}$ did in Eq. (\ref{e:perc}).  We need to study separately monomers,
``matched'' dimers (in which both nodes in the dimer survive the
initial attack), and ``unmatched'' dimers, (in which only one of the two
nodes survives). Eq. (\ref{e:gen}) is a general framework which must be
adapted to account for the three different types of nodes in our system. We
thus obtain three equations from Eq. (\ref{e:gen}). The equation for $a_m$ studies links that leave monomers, while the equation for $a_d$ studies links that leave matched dimers and the equation for
$a_u$ studies links that leave unmatched dimers. We then calculate the probability that $A_j$ (the node to which the link leads) is a given type of node (for example, a monomer), and then multiply this by the probability that if $A_j$ is a monomer (for example), $A_j$ does not connects $A_i$ to the giant component, given our {\it a priori} assumption. Once we calculate these probabilities for each of the three types of nodes, we add these probabilities, finally arriving at the total probability that a node does not lead to the giant  component, given our {\it a priori} assumption that its support node does. We find
\begin{widetext}
\begin{align}
a_d=&(1-p)+qp(a_m^{k-1})+&(1-q)p(1-p)&(a_u^{k-2}+[1-a_u^{k-2}]a_u^{k-1})&+(1-q)&p^2(a_d^{2k-3})
\label{e:ad}\\
a_m=&(1-p)+&p(1-p)&(a_u^{k-2}+[1-a_u^{k-2}]a_u^{k-1})&+&p^2(a_d^{2k-3}).
\label{e:am}
\end{align}
\end{widetext}
These two equations are relatively simple; $P(x\cup y)$ is either just $P(x)$ or $P(x)+P(y)-P(x\cap y)$. That is, $y$, failure due to the second network, is either 1) impossible due to the \textit{a priori} assumption, or 2) independent of $x$. Case 1 occurs when both $A_i$ and $A_j$ are monomers or parts of matched dimers; in these cases, nodes that survived the initial attack connect $D(A_j)$ and $D(A_i)$, allowing $D(A_j)$ to be connected to the giant component. Case 2 occurs when $A_j$ is a part of an unmatched dimer; due to the local tree-like structure of the large RR network, the path that connects $A_j$ to its giant component will not overlap with the path that connects $D(A_j)$ to its giant component. Thus, the existence of one path has no correlation with the existence of the other. 
The calculation of $a_u$ is more difficult. The events $x$ and $y$ are not always independent because
$A_j$'s and $B_j$'s paths to the mutual giant component may overlap. We must find $P(x\cap
y)\neq P(x)P(y)$. Therefore, we introduce another two variables, $z_m$
and $z_d$, which denote the probabilities that neither the link from $A_i$ to $A_j$ nor the link from $B_i$ to $B_j$ lead to the giant component. The principle in these cases is the same as the one used in the determination of $a_m$ and $a_d$. $P(x\cap y)$ can be expressed as a power of $z_m$ or $z_d$. We find:
\begin{widetext}
\begin{align}
a_u=&(1-p)+qp(2a_m^{k-1}-z_m^{k-1})+&(1-q)p(1-p)(a_u^{k-2}+[1-a_u^{k-2}]a_u^{k-1})+&(1-q)p^2(2a_d^{2k-3}-z_d^{2k-3})
\label{e:au}\\
z_d=&q(1-p)+(1-q)(1-p)^2+qp(z_m^{k-1})+&2(1-q)p(1-p)(a_u^{k-2}+[1-a_u^{k-2}]a_u^{k-1})+&(1-q)p^2(z_d^{2k-3})
\label{e:zd}\\
z_m=&(1-p)^2 +&2p(1-p)(a_u^{k-2}+[1-a_u^{k-2}]a_u^{k-1})+&p^2(z_d^{2k-3}).
\label{e:zm}
\end{align}
\end{widetext}
Finally, the fraction of nodes in the mutual giant component is 
\begin{multline}
\mu=p\{1-m(2a_m^k-z_m^k)\\
-(1-m)[p(2a_d^{2k-2}-z_d^{2k-2})+(1-p)(2a_u^{k-1}-a_u^{2k-2})]\}.
\label{e:mu}
\end{multline}
To solve the equations (Eqs. (\ref{e:ad})-(\ref{e:zm})), we define
$\vec{\xi}\equiv(a_d,a_u,a_m,z_d,z_m)$, and a function $\vec{\varphi}(\vec{\xi})$, which represents the right hand sides of Eqs. (\ref{e:ad})-(\ref{e:zm}). These
equations always have a trivial solution of $(1,1,1,1,1)$, for which $\mu=0$. Through an
iterative process $\vec{\xi}_0=\vec{0}$,
$\vec{\xi}_n=\vec{\varphi}(\vec{\xi}_{n-1})$, we search for a
non-trivial fixed point $\vec{\xi}=\vec{\varphi}(\vec{\xi})$ for different values of $p$. The results for $\mu(p)$ are presented in 
Figure \ref{f:solve}. Our results for $p_c$
coincide with our numerical data to within the precision of the simulation. We
remarkably see three types of transitions in the value of $\vec{\xi}$
as a function of $p$, as the simulations indicated. For $k\leq 7$, we observe a
second-order transition. For $k\geq 9$, we find a first-order
transition. For $k=8$, as $p$ decreases, we first see a first-order transition at
$p=p_{cI}=.2762$, where $\mu$ changes dramatically (but not to zero) and a second-order transition to $\mu=0$ when $p$ decreases to $p_{cII}=.2688$.\\ 

In conclusion, our results demonstrate the interesting behavior of collapsing random networks with distance-restricted dependency links.
We find that networks with long dependency distances, 
are much more vulnerable than networks with short dependency distances.
Moreover, the networks with large dependency distance and large degree
collapse via an abrupt first-order transition, near which a removal of a single additional node can create a collapse of the entire system, while
the networks with short dependency distances and low degree gradually disintegrate via a continuous second-order transition without catastrophic cascades, which make such networks safer.
For the transitional case, we have found a two-stage
collapse. The first-order transition is sudden, as
expected, but the network survives until the second transition, which completely destroys the networks.  Additionally, we have developed a new method
to find the critical value of the scope of the initial attack, $1-p_c$, which will likely prove useful in future studies
of interdependent networks that are locally similar to each other.
\\
We wish to thank DTRA for financial support and Shlomo Havlin for stimulating discussions.  
We acknowledge the partial support of this research
through the Dr. Bernard W. Gamson Computational Science Center at Yeshiva
College.

\end{document}